\documentclass[aps,prd,twocolumn,amsfonts]{revtex4}
\usepackage{amsmath}
\usepackage{graphicx}
\def\D{{\cal D}}
\def\R{\,^{(5)\!} R}
\def\G{\,^{(5)\!} G}
\def\g{\,^{(5)\!}g}

\def\vu{^{(5)\!}}
\newcommand{\be}{\begin{equation}}
\newcommand{\ee}{\end{equation}}
\newcommand{\bea}{\begin{eqnarray}}
\newcommand{\eea}{\end{eqnarray}}
\begin{document}

\title{Gravitational waves from brane-world inflation
with induced gravity}

\author{Mariam Bouhmadi-L\'opez}\email{mariam.bouhmadi@port.ac.uk}

\author{Roy Maartens}\email{roy.maartens@port.ac.uk}

\author{David Wands}\email{david.wands@port.ac.uk}

\affiliation{\vspace*{0.2cm} Institute of Cosmology \&
Gravitation, University of Portsmouth, Portsmouth~PO1~2EG, UK
\vspace*{0.2cm}}

\date{\today}

\begin{abstract}

We calculate the amplitude of gravitational waves produced by
inflation on a de Sitter brane embedded in five-dimensional
anti-de Sitter bulk spacetime, extending previous calculations in
Randall-Sundrum type cosmology to include the effect of induced
gravity corrections on the brane. These corrections arise via a
term in the brane action that is proportional to the brane Ricci
scalar. We find that, as in the Randall-Sundrum case, there is a
mass gap between the discrete zero-mode and a continuum of massive
bulk modes, which are too heavy to be excited during inflation. We
give the normalization of the zero-mode as a function of the
Hubble rate on the brane and are thus able to calculate the high
energy correction to the spectrum of gravitational wave (tensor)
modes excited on large scales during inflation from initial vacuum
fluctuations on small scales. We also calculate the amplitude of
density (scalar) perturbations expected due to inflaton
fluctuations on the brane, and show that the usual
four-dimensional consistency relation for the tensor/scalar ratio
remains valid for brane inflation with induced gravity
corrections.

\end{abstract}


\maketitle

\section{Introduction}

String theory and M~theory have motivated interest in brane-world
models that describe extra-dimensional gravity. In these models,
the observable four-dimensional (4D) universe is a brane
hypersurface embedded in a higher-dimensional bulk space. A simple
but rich phenomenology for cosmology is provided by the
Randall-Sundrum (RS) models with a single brane in a
five-dimensional (5D) anti de Sitter (AdS$_5$) spacetime. (For
recent reviews, see Ref.~\cite{review}.) Although the fifth
dimension is infinite, the zero-mode of the 5D graviton,
corresponding to 4D gravitational waves, is localized at low
energies on the brane due to the warped geometry of the bulk. This
allows one to recover general relativity in the low-energy limit.

Quantum corrections to the RS model are expected to arise from the
induced coupling of brane matter to bulk gravitons. This is the
so-called induced gravity (IG) effect, that leads to the
appearance of terms proportional to the 4D Ricci scalar in the
brane action~\cite{ch,dgp,others}. In the Dvali, Gabadadze and
Porrati (DGP) model~\cite{dgp} the induced gravity brane-world was
put forward as an alternative to the RS one-brane model, in which
general relativity was recovered, also despite an infinite extra
dimension, but without warping in 5D Minkowski spacetime (the
brane had no tension). In contrast to the RS case with high-energy
modifications to general relativity, the DGP brane-world produced
a low-energy modification, leading to late-time acceleration even
in the absence of dark energy~\cite{deff}.

Here we will take a different viewpoint: we are mainly interested
in the IG effect as a correction to the RS brane-world at high
energies, rather than as a low-energy alternative (see also
Refs.~\cite{ktt,mmt,pz,zc,bw}). In this case the IG effect will
operate at high energies, above the brane tension, which is the
threshold energy for RS modifications to general relativity. As a
consequence there is no late-time IG modification to general
relativity, but rather an early-time modification to the RS model.
High-energy inflation will be subject to the IG effect, and we
seek to determine in particular the consequent changes to the
gravitational waves generated during inflation. If the IG
correction becomes large so that it is dominant, then there is no
high-energy RS regime in the early universe, and we recover the
original DGP model.

During slow-roll inflation, zero-point quantum fluctuations of
light fields (whose effective mass is less than the Hubble scale)
are swept up to super-Hubble scales and acquire a nearly
scale-invariant spectrum. In the RS brane-world, the zero-mode of
the 5D graviton is a massless spin-2 field on the brane that
represents 4D gravitational waves. But there are also massive
modes on the brane that arise via the projection onto the brane of
the 5D graviton. These massive modes could qualitatively alter the
spectrum of gravitational waves generated during inflation. It
turns out that there is a mass gap between the zero-mode and the
the massive modes~\cite{gs}, so that the massive modes are too
heavy to be excited and the spectrum is qualitatively the same as
in general relativity. However, the amplitude of the zero-mode is
boosted at high energies, compared with the standard 4D
case~\cite{lmw}.

We show that significant changes to the RS case are introduced by
the IG term at high energies: the IG correction acts to limit the
growth of amplitude, relative to the 4D case, as the energy scale
of inflation grows. Then we investigate the scalar perturbations
produced in inflation (assuming there is no contribution from 5D
gravitons), and show that the scalar amplitude has the same
qualitative behaviour as the tensor amplitude. We also show that
the standard tensor-to-scalar consistency relation continues to
hold in the presence of the IG correction.

\section{Field Equations}

We assume that the 5D bulk contains only negative vacuum energy,
and the gravitational field obeys Einstein gravity. The 4D brane
has a positive brane tension, and an induced gravity term
localized on it. Then the gravitational action is
 \bea
{\cal S} &=& \frac{1}{2\kappa_5^2} \int d^5x \sqrt{-\,\vu g}
\left[ {\R}-2\Lambda_5 \right] \nonumber\\&&~{} + \int_{\rm brane}
d^4x\, \sqrt{-g}\left[- \sigma + \frac{\gamma}{2\kappa_4^2}R +
{\cal L}_{\rm matter} \right]\!, \label{action}
 \eea
where $\gamma$ is a dimensionless constant controlling the
strength of the IG correction, with $\gamma=0$ giving the RS
model. The induced metric on the brane is $g_{ab}=\,\vu
g_{ab}-n_an_b$, where $n^a$ is the unit normal, and $\sigma\,
(\geq 0)$ is the brane tension, while $\Lambda_5\,(\leq 0)$ is the
bulk cosmological constant. The fundamental energy scale of
gravity is the 5D scale $M_5$, where $\kappa_5^2=8\pi/M_5^3$. The
4D Planck scale $M_4\sim 10^{19}~$GeV is an effective scale, fixed
by the effective gravitational coupling constant $\kappa_4^2=8\pi
G=8\pi/M_4^2$ on the brane at low energies (typically $M_4 \gg
M_5$). Note that for the low- and high-energy limits to yield a
positive effective gravitational coupling constant, we require
$0\leq\gamma<1$.

The 5D field equations following from the bulk action are
unchanged from the RS case:
\begin{eqnarray}
\G_{ab} &=& -\Lambda_5 \g_{ab}\,.\label{5d}
\end{eqnarray}
The IG correction term enters via the junction conditions at the
brane. Assuming mirror ($\mathbb{Z}_2$) symmetry about the brane,
these are~\cite{mmt}
\begin{eqnarray}
K_{\mu\nu}-Kg_{\mu\nu} &=& -{\kappa_5^2\over2}
\left(T_{\mu\nu}-\sigma g_{\mu\nu}-{\gamma \over
\kappa_4^2}G_{\mu\nu} \right) \,,\label{gjc}
\end{eqnarray}
where $K_{\mu\nu}$ is the extrinsic curvature and $T_{\mu\nu}$ is
the energy-momentum tensor of matter on the brane, which obeys the
conservation equations~\cite{mmt}
\begin{equation}
\label{tmunu} \nabla^\nu T_{\mu\nu}=0\,.
\end{equation}

The length scale $\ell$ and energy scale $\mu$ associated with the
bulk curvature are given by
\begin{eqnarray}
\Lambda_5 &=& -{6\over \ell^2}=- 6\mu^2\,.\label{lam}
\end{eqnarray}
If we impose the RS fine-tuning for the brane tension,
$\kappa_5^2\sigma=\sqrt{-6\Lambda_6}$, this sets the brane
cosmological constant to zero, and then we have~\cite{mmt,bw}
\begin{equation}
\kappa_4^2 \equiv \kappa_5^2\mu(1-\gamma) \,. \label{kappa}
\end{equation}
The general form of the effective field equations projected on the
brane is given in Ref.~\cite{mmt}.

We will be interested in 4D tensor metric perturbations about a de
Sitter brane (modelling extreme slow-roll inflation) in an AdS$_5$
bulk. Such a brane is a solution of the junction
equations~(\ref{gjc}) with a constant energy density on the brane,
$\rho>0$, and hence a constant Hubble rate on the brane, $H$. The
background space-time metric, i.e., an AdS$_5$ bulk sliced into 4D
slices with de Sitter geometry, may be written in the Gaussian
normal form
\begin{equation}
ds^2=n^2(y)\left[-dt^2+\exp(2H_{\epsilon}t)
d\vec{x}^2\right]+dy^{2}. \label{background}
\end{equation}
The warp factor in the bulk is
\begin{eqnarray}
n(y)&=&\frac{H_{\epsilon}}{\mu}\sinh
\left[\mu\left(y_{\epsilon}+\epsilon|y|\right)\right] .
\label{n(y)}
\end{eqnarray}
The brane is fixed at $y=0$ and we choose $n(0)=1$ on the brane,
so that $y_\epsilon$ is given by
\begin{eqnarray}
y_{\epsilon}&=&\frac{1}{\mu}\,\mbox{arcsinh}
\,\frac{\mu}{\,\,H_{\epsilon}}\,.\label{yepsilon}
\end{eqnarray}
The Hubble parameter $H_\epsilon$ is given by the modified
Friedman equation
\begin{eqnarray}
&& \! H^2_{\epsilon}=\frac{\kappa_4^2}{3\gamma}\Biggl\{\rho  +
\sigma
\nonumber\\
&&\!{} + \frac{6\kappa_4^2}{\gamma\kappa_5^4}\! \left[1 +
\epsilon\sqrt{1+\frac{\gamma\kappa_5^4}{3\kappa_4^2}\left(\rho+\sigma
+ \frac{\gamma\sigma^2\kappa_5^4}{12\kappa_4^2}\right)}\,\,\right]
\Biggr\}\!\!, \label{Friedmann2}
\end{eqnarray}
where we have imposed Eq.~(\ref{kappa}). In the
expressions~(\ref{background})--(\ref{Friedmann2}), $\epsilon=\pm
1$ corresponds to the choice of sign for the square-root in
Eq.~(\ref{Friedmann2}). We will refer to $\epsilon=+1$ as the
positive branch and $\epsilon=-1$ as the negative branch. In the
following we take the energy density, $\rho$, to be constant,
corresponding to the inflaton energy density in the extreme slow
roll limit. We recover the Randall-Sundrum case~\cite{lmw} in the
limit $\gamma\rightarrow 0$, but only on the negative branch.

On the positive branch, the bulk coordinate $y$ can take any real
value and the warp factor $n(y)$ reaches its minimum value at the
brane. On the other hand, in the negative branch, the bulk
coordinate is bounded such that $|y|\leq y_-$, where $|y|=y_-$
corresponds to the location of the Cauchy horizon, where
$n(y_-)=0$. In this case, the warp factor has its maximum at the
brane. We recover the original RS model, with a Minkowski brane,
as $H\to0$ and $y_-\to\infty$ for $\rho\to0$, on the negative
branch. The original DGP model with $H_+>0$ is obtained in the low
energy limit on the positive branch as $\rho+\sigma\to0$.

For later convenience, we introduce the conformal bulk coordinate,
with $dz=dy/n(y)$,
\begin{eqnarray}
z(y)=-\frac{\epsilon}{\mu} \mbox{sgn}(y)\sinh \mu
y_{\epsilon}\,\ln\,\coth\left[\frac{\mu}{2}\left(y_{\epsilon}+\epsilon
|y| \right)\right]\!.
\end{eqnarray}
For the positive branch $0<|z|\leq z_{b+}$, where $z_{b+}$
corresponds to the location of the brane, $z_{b+}=|z(0)|$. When
$|z|\rightarrow 0$, the bulk coordinate $y$ diverges. For the
negative branch, the conformal bulk coordinate is unbounded, $|z|
\geq z_{b-} $, where $z_{b-}$ gives the location of the brane. In
this case, the Cauchy horizons are located at $|z|\rightarrow
+\infty$. Finally, the warp factor becomes
\begin{eqnarray}
n(z)=\frac{H_{\epsilon}}{\mu\sinh H_{\epsilon}|z|}\,. \label{n(z)}
\end{eqnarray}

\section{Bulk metric perturbations}

We now consider tensor metric perturbations (from the viewpoint of
a 4D observer), $\vu {g}_{ab}\to \,\vu {g}_{ab}+\delta \,\vu
{g}_{ab}$. The perturbed metric is~\cite{lmw}
\begin{equation}
ds^2 = n^2(y)\left[ -dt^2 + e^{-2H_{\epsilon}t}\!
\left(\delta_{ij} + h_{ij}\right) \,dx^idx^j \right] + dy^2\!,
\end{equation}
where $h_{ij}$ is transverse and traceless. The wave equation in
the bulk for the perturbations is
\begin{equation}
\delta\G^a{}_b=0\,,
\end{equation}
the same as in the RS case. This means that the bulk mode
solutions for metric perturbations will be the same as in the RS
case~\cite{lmw}, but the IG junction conditions will introduce
changes to the normalization and amplitudes of the modes, as
discussed below.

We decompose the general metric perturbation into Fourier modes
\begin{equation}
h_{ij}(t,\vec{x},y)=E(t,y)e_{ij}(\vec{x}),
\end{equation}
where $e_{ij}$ is a transverse traceless harmonic on the spatially
flat 3-space, i.e., $\vec{\nabla}^2 e_{ij}=-k^2 e_{ij}$ . Then the
equation of motion for $E$ is~\cite{lmw}
\begin{equation}
\ddot{E}+3H_{\epsilon}\dot{E} +
k^2\,e^{-2H_{\epsilon}t}E-n^2\left(E''+4\frac{n'}{n}E'\right) =
0\,, \label{Eequation}
\end{equation}
where a dot denotes a derivative with respect to $t$ and a prime a
derivative with respect to $y$. The perturbation $E$ may be
decomposed into Kaluza-Klein (KK) modes
\begin{equation}
E(t,y) = \int dm\,\psi_{m}(t)\,\mathcal{E}_m(y)\,,
\end{equation}
and the wave equation~(\ref{Eequation}) then separates to give
\begin{eqnarray}
\ddot{\psi}_m+ 3H_{\epsilon}\dot\psi_m+
\left[k^2\,e^{-2H_{\epsilon}t}+m^2\right] \psi_m=0\,,\\
\label{StLi} (n^4\mathcal{E}_m')'+m^2 n^2 \mathcal{E}_m=0\,.
\end{eqnarray}

Although the massive modes of the tensor perturbations satisfy the
same bulk wave equation as in the RS case~\cite{lmw}, the junction
condition at the brane is very different. The tensor part of
Eq.~(\ref{gjc}) implies
\begin{equation}
\delta K^{\mu}{}_\nu=\frac{\gamma\kappa_5^2}{2\kappa_4^2} \,\delta
G^\mu{}_\nu =\frac{\gamma}{2\mu(1-\gamma)} \,\delta G^\mu{}_\nu
\,, \label{junc1}
\end{equation}
where we have neglected any tensor contributions of the
anisotropic stress exerted by matter on the brane, and the second
equality follows from Eq.~(\ref{kappa}). In the RS case
($\gamma\rightarrow 0$), we recover the Neumann boundary
condition. On the brane, the non-vanishing components of the
Einstein tensor are given by~\cite{lmw,Bridgman:2001mc}
\begin{equation}
\delta G^i{}_j=\frac12\left[\ddot{h}^i{}_j + 3H_{\epsilon}
\dot{h}^i{}_j-e^{-2H_{\epsilon}t}\vec{\nabla}^2\, h^i{}_j\right]
\,,
\end{equation}
and consequently, the boundary condition for $\mathcal{E}_m$ is
\begin{equation}
\mathcal{E}_m'(0)=-\frac{m^2\gamma}{2\mu(1-\gamma)}
\,\mathcal{E}_m(0)\,. \label{jc}
\end{equation}
It is important to note that this boundary condition depends on
the mass of the modes, $m^2$, due to the IG term. Only the
zero-mode, $m=0$, has the same boundary condition as in the RS
case. As a result of this new feature, the scalar product of the
eigenmode functional space has to include suitable boundary terms.
This is qualitatively the same as the case of Gauss-Bonnet
modifications to the RS model~\cite{cdd,GB}.

In what follows we show that the eigenmodes resulting from
Eqs.~(\ref{StLi}) and (\ref{jc}) are orthonormal with respect to
the following scalar product:
\begin{eqnarray}
(\mathcal{E}_m , \mathcal{E}_{\tilde m}) &=& 2
\int_{0}^{U_\epsilon}\, dy \, n^2
\mathcal{E}_m\,\mathcal{E}_{\tilde m}
+\frac{\gamma}{\mu(1-\gamma)}
\,\mathcal{E}_m(0) \,\mathcal{E}_{\tilde m}(0)\nonumber\\
&=& \delta(m,{\tilde m}) \,, \label{orthonorm1}
\end{eqnarray}
where $U_+=\infty, U_-=y_-$, and $\delta(m,\tilde{m})$ denotes a
Kronecker symbol for the discrete modes and a Dirac distribution
for the continuous ones. For $\gamma \rightarrow 0 $ the scalar
product reduces to the one used in the RS case.

\section{Kaluza-Klein modes}

We consider a single de Sitter brane embedded in AdS space so that
we only impose one boundary condition, Eq.~(\ref{jc}). One could
also consider the presence of a second
brane~\cite{2brane,padilla}, with a corresponding second boundary
condition, which would yield a discrete spectrum of bulk modes.

Our discussion of the spectrum of bulk modes will follow closely
that of Ref.~\cite{GB}. Defining $\phi_m\equiv
n^{3/2}\mathcal{E}_m$, we rewrite the wave equation~(\ref{StLi})
in a way that incorporates the junction condition~(\ref{jc}):
\begin{equation}  \label{zwe}
-\D_{(+)} \left[ q(z) \, \D_{(-)} \, \phi_m(z)\right] = m^2 \,
w(z) \, \phi_m(z)\,,
\end{equation}
where
\begin{equation}\label{D}
\D_{(\pm)}=\frac{d}{dz} \pm
\frac{3}{2}\,\frac{1}{n}\frac{dn}{dz}\,,
\end{equation}
and
\begin{eqnarray}  \label{q}
q&=&\theta\left[\epsilon(z_{b\epsilon}-z)\right]
-\epsilon\theta\left[-z-z_{b\epsilon}\right]\,,\\
\label{w} w &=&q(z)+\,\frac{\gamma}{\mu(1-\gamma)}\,
\delta(z+\epsilon z_{b\epsilon})\,.
\end{eqnarray}
Here $\theta$ is the heaviside function, and the $\pm$ subscripts
in $\D_\pm$ are not to be confused with the branch parameter
$\epsilon$.

From Eqs.~(\ref{n(z)}), (\ref{q}) and (\ref{w}) it follows that
for $| z |\neq z_{b\epsilon}$, we have $q=w=1$. Thus for $|z|\neq
z_{b\epsilon}$, Eq.~(\ref{zwe}) reduces to the Schr\"odinger-type
equation,
\begin{eqnarray}
-\D_{(+)}\D_{(-)}\phi_m &=& -\frac{d^2\phi_m}{dz^2} +
\bigg[\frac{15}{4}\, \frac{H_{\epsilon}^2}{\sinh^2 H_{\epsilon}z}+
\frac{9}{4}H_{\epsilon}^2\bigg]\phi_m \nonumber\\
{}&=&m^2\,\phi_m \,, \label{schrod}
\end{eqnarray}
exactly as in the RS case ($\gamma\rightarrow 0$)~\cite{gs,lmw}.

The boundary condition, Eq.~(\ref{jc}), for $\phi_m$ at the brane
is
\begin{equation}
\label{jc2} \D_{(-)}\;\phi_m(-\epsilon z_{b\epsilon})= -m^2
\frac{\gamma}{2\mu(1-\gamma)} \phi_m(z_{b\epsilon})\,.
\end{equation}
(Note that in the negative branch, $y\rightarrow 0^+$ implies
$z\rightarrow z_{b-}^+$, while for the positive branch,
$y\rightarrow 0^+$ implies $z\rightarrow -z_{b+}^+$.)
Equation~(\ref{jc2}) may also be obtained by matching the
distributional parts of Eq.~(\ref{zwe}). It may be checked that
the eigenmodes resulting from Eqs.~(\ref{schrod}) and (\ref{jc2})
are orthonormal with respect to the scalar product
Eq.~(\ref{orthonorm1}), which results from
\begin{equation}
(\phi_m , \phi_{\tilde m}) =\int_{\mathrm{bulk}} dz \,w \phi_m
\phi_{\tilde m}\,, \label{orthonorm2}
\end{equation}
when $\mathbb{Z}_2$-symmetry is imposed and the distributional
term in Eq.~(\ref{w}) is taken into account.

We now investigate the spectrum of modes resulting from
Eqs.~(\ref{schrod}) and (\ref{jc2}). There is the same zero-mode
solution as in the RS case~\cite{lmw}
\begin{equation}
\label{zero}
\phi_0(z)=C_{\epsilon} \,n^{3/2}(z)\,,
\end{equation}
where $\mathcal{E}_0=C_{\epsilon}$ is a constant. This zero-mode
solution obeys the boundary condition (\ref{jc2}) for both the
positive and negative branches in Eq.~(\ref{Friedmann2}).

The zero mode in the negative branch is normalizable as in the RS
case. Using Eqs.~(\ref{orthonorm2}) and (\ref{zero}), the
condition $(\phi_0,\phi_0)=1$ on the negative branch gives the
normalization constant $C=C(H_-/\mu)$ as a function of the Hubble
rate relative to the AdS scale, $\mu$, where
\begin{eqnarray}
\label{psi0} C_-^{-2}(x)
 =
\frac{\gamma}{\mu(1-\gamma)} + \frac{1}{\mu} \left[
\sqrt{1+x^2}-x^2\mbox{arcsinh}\,\frac{1}{x}\right]
 \!. \nonumber\\
\end{eqnarray}
This reduces to the RS result~\cite{lmw} when $\gamma=0$.

However, the zero-mode is not normalizable in the positive branch
(for a single brane), since $\phi_0$ diverges as $z\to0$. We will
henceforth restrict the analysis to the negative branch, and drop
the negative subscript, i.e., $H_-\equiv H, z_{b-}\equiv z_b$.

As in the RS case~\cite{lmw,gs}, the potential in the
Schr\"odinger equation (\ref{schrod}) approaches $\frac{9}{4}H^2$,
as $|z|\to\infty$, which gives the threshold between light modes,
with $m^2<{9\over4}H^2$, and heavy modes, with
$m^2>\frac{9}{4}H^2$. For heavy modes, the two linearly
independent solutions of Eq.~(\ref{schrod}) oscillate with
constant amplitude as $|z|\rightarrow \infty$. The boundary
condition Eq.~(\ref{jc2}) gives $\phi_m(z)$ as a particular
combination of these two solutions, for every $m$. Thus we have a
continuous spectrum of normalizable modes with $m^2>
{9\over4}H^2$.

For light modes, the general solution of the Schr\"odinger
equation (\ref{schrod}) contains an asymptotically growing
exponential and a decaying exponential as $|z|\to\infty$. Thus we
would expect only a discrete spectrum of normalizable modes
corresponding to particular values of $m^2\leq {9\over 4}H^2$ for
which the junction condition (\ref{jc2}) selects only the
asymptotic decaying mode. This is exactly what happens to give the
normalizable zero-mode (\ref{zero}) for the negative branch in
Eq.~(\ref{Friedmann2}).

In order to see whether the junction conditions allow for
normalizable light modes other than the zero-mode, we introduce
new mode functions, following Ref.~\cite{GB},
\begin{equation}
\label{Fi} \Phi_m(z)=\D_{(-)}\;\phi_m(z)\,.
\end{equation}
These are the partners of the modes $\phi_m$ in super-symmetric
quantum mechanics~\cite{cks}, and have the same spectrum except
for the zero-mode, $\Phi_0$, which vanishes identically, by
Eqs.~(\ref{D}) and (\ref{zero}). Acting on Eq.~(\ref{schrod}) with
the operator $\D_{(-)}$, we obtain the wave equation for $\Phi_m$,
\begin{eqnarray}
-\D_{(-)}\D_{(+)}\Phi_m &=& -\frac{d^2\Phi_m}{dz^2} +
\bigg[\frac{3}{4}\,\frac{H^2}{\sinh^2 Hz}+\frac{9}{4}H^2\bigg]
\Phi_m \nonumber\\ &=& m^2\,\Phi_m \,. \label{schrod2}
\end{eqnarray}
The junction condition for $\Phi_m$ follows from
Eqs.~(\ref{schrod}) and (\ref{jc2}),
\begin{equation}
\frac{d\Phi_m}{dz}(z_{b})=\left[\frac32\sqrt{H^2+\mu^2}\,
+\frac{2\mu(1-\gamma)}{\gamma}\right] \Phi_m(z_{b})\,, \label{jc3}
\end{equation}
for $\gamma\neq0$, while Eqs.~(\ref{jc2}) and (\ref{Fi}) show that
$\Phi_m(z_{b})=0$ for $\gamma=0$. The important simplification
here is that the boundary condition no longer involves the mass of
the modes, and reduces to Dirichlet-type for $\gamma=0$.
Multiplying Eq.~(\ref{schrod2}) by $\Phi_m$ and integrating by
parts, we find that
\begin{eqnarray}
&&\left(m^2-\frac{9}{4}H^2\right)\int_{z_{b}}^{\infty}\! dz\,
\Phi_m^2 = \frac{3}{4}H^2\int_{z_{b}}^{\infty}\! dz\,
\frac{\Phi_m^2} {\sinh^2Hz}\nonumber\\ &&~~~{}
 +\int_{z_{b}}^{\infty}\! dz\,\Phi_m^{'2} +
\bigg[-\Phi_m\Phi_m'\bigg]_{z_{b}}^{\infty}. \label{proof}
\end{eqnarray}
Any normalizable light mode, $\phi_m$, with $m^2<{9\over4}H^2$,
must decrease exponentially as $|z|\rightarrow \infty$, and so
must its partner, $\Phi_m$. The corresponding upper boundary term
at infinity in Eq.~(\ref{proof}) therefore vanishes. The lower
boundary term on the brane, according to Eq.~(\ref{jc3}), is
positive, remembering that we have restricted our analysis to
$0\leq\gamma< 1$. In this case, the right-hand side of
Eq.~(\ref{proof}) is positive, while the left-hand side is
negative, except for $\Phi_m=0$, i.e., for $m=0$. We can therefore
conclude that for $0\leq\gamma<1$, there are no normalizable light
modes in the mass gap. The spectrum of KK modes in the negative
branch of Eq.~(\ref{Friedmann2}) is the same as the RS case, i.e.,
it consists of the massless bound-state zero-mode, and a continuum
of states with $m>{3\over2}H$.

As in the RS case ($\gamma\rightarrow 0$)~\cite{gs,lmw} and as in
the case of Gauss-Bonnet gravity in the bulk~\cite{GB}, we have
only one discrete light mode with $m^2=0$ and a then mass-gap
before the continuum of heavy modes. On the other hand, we have
shown that there is no zero-mode for the positive branch,
$\epsilon=+1$, and it remains to be seen whether or not there are
light modes with $m^2<{9\over4}H^2$ in this case.

\section{Quantum fluctuations on the brane}

Having identified the normalizable bulk modes for a single de
Sitter brane in an anti-de Sitter bulk, we will now estimate the
spectrum of graviton fluctuations generated in de Sitter inflation
on the brane for the negative branch in Eq.~(\ref{Friedmann2}). We
treat each normalizable mode as a quantum field in four
dimensions, as in the RS case~\cite{lmw,fk}. Taking these 4D
fields to be in an initial vacuum state on small scales is
consistent with taking an incoming AdS vacuum state in the
five-dimensional viewpoint~\cite{grs}.

Massive modes with $m^2>{9\over4}H^2$ remain underdamped even on
large scales, and fluctuations are strongly suppressed on
super-horizon scales. They remain in their vacuum
state~\cite{lmw,fk}.

However initial vacuum fluctuations in the zero-mode become
over-damped as they are stretched beyond the Hubble scale. The
zero-mode thus acquires a spectrum of classical perturbations on
super-horizon scales. For $m^2=0$, the effective action has the
standard form of 4D general relativity, except for the overall
factor $\kappa_5^2$ instead of $\kappa_4^2$~\cite{GB,padilla},
which rescales the amplitude of quantum fluctuations
accordingly~\cite{lmw,GB}.  The amplitude of the zero-mode metric
fluctuations on the brane, \mbox{$\phi_0(z_{b})=C(H/\mu)$}, where
$C(H/\mu)$ is given in Eq.~(\ref{psi0}), then introduces a further
rescaling relative to the 4D result, which is dependent on the
Hubble scale relative to the AdS scale. Following the notation of
Ref.~\cite{llkcba}, the amplitude of gravitational waves produced
on super-horizon scales on the brane is thus given by
\begin{equation}
\label{tensoramplitude} A_T^2 = \frac{2\kappa_4^2}{25} \;
\bigg(\frac{H}{2\pi}\bigg)^2 \; F_\gamma^2(H/\mu) \,,
\end{equation}
where the correction to standard 4D general relativity is given by
\begin{equation}
F_{\gamma}^2 (H/\mu) = \frac{\kappa_5^2}{\kappa_4^2} C^2(H/\mu)
\,.
\end{equation}
From Eq.~(\ref{psi0}), we have
\begin{eqnarray}
 \!\! F_{\gamma}^{-2}(x) =\gamma
+(1-\gamma)\! \left[ \sqrt{1+x^2} - x^2\mbox{arcsinh}\,\frac{1}{x}
\right]\!. \label{F}
\end{eqnarray}
This correction depends on the IG coupling (through the parameter
$\gamma$) and on the energy scale at which inflation occurs,
relative to the 5D curvature scale $\mu$. It reduces to the result
of Ref.~\cite{lmw} for the RS case ($\gamma\rightarrow 0$).

When $x \equiv H/ \mu\to 0$, we have $F_{\gamma}\to 1$ and we
recover the standard 4D result~\cite{llkcba}. The amplitude of the
normalized zero-mode on the brane gives the ratio between the
effective 4D Newton constant at low energies, $\kappa^2_4$, and
the 5D constant, $\kappa^2_5$.

\begin{figure}[h]
\begin{center}
\hspace*{-1cm}\includegraphics[width=9cm]{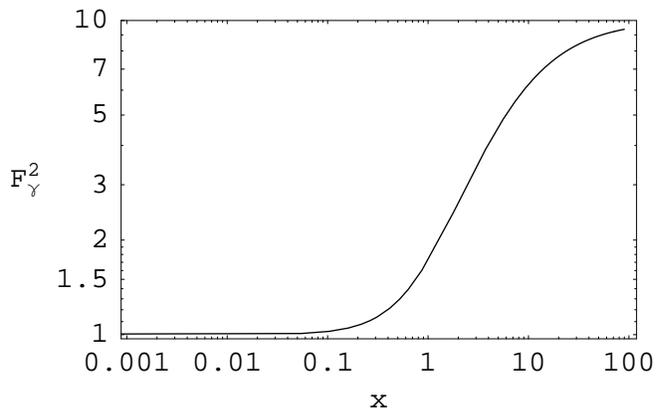}\\
\caption{The dimensionless amplitude $F_{\gamma}^2$ of the tensor
zero-mode relative to the 4D general relativity result, plotted
against the dimensionless energy scale of inflation, $H/\mu$, for
$\gamma=0.1$.}\label{F1}
\end{center}
\end{figure}

For $0<\gamma<1$ we find that tensor fluctuations are enhanced
relative to the 4D result ($F_\gamma>1$) at high energies.
However, unlike the RS case, the amplitude of the tensor zero-mode
relative to the 4D general relativity result is bounded, $1\leq
F_{\gamma}^2<1/\gamma$. For $H\gg \mu$ and $0<\gamma<1$, we have
\begin{equation}
\label{Falpha} F_{\gamma}^2(H/\mu) \approx \frac{1}{\gamma},
\end{equation}
while the RS case ($\gamma=0$) yields~\cite{lmw}
\begin{equation}
\label{Fzero}F_0^2(H/\mu) \approx {3\over2}\,\frac{H}{\mu}\,.
\end{equation}
The qualitative behaviour is illustrated in Fig.~\ref{F1} for
$\gamma=0.1$.

\section{The tensor to scalar ratio}

The contribution of tensor perturbations to cosmic microwave
background (CMB) anisotropies is an important quantity for
constraining inflationary models~\cite{liddle}. In the RS case,
both tensor and scalar perturbations on the brane are enhanced
relative to a 4D inflation model at a fixed energy scale, but the
tensors are enhanced less and thus the relative tensor
contribution is suppressed in comparison with the 4D
case~\cite{mwbh}. In this section we compute the scalar
perturbation amplitude $A_S$ including the effect of the IG term..

In standard, slow-roll inflationary models driven by a single
inflaton field there is a consistency relation between the
contribution to the CMB from scalar (density) perturbations and
tensor modes (gravitational waves). (For a review, see, e.g.,
Ref.~\cite{llkcba}). To lowest order in the slow-roll
approximation, the ratio of the tensor to scalar perturbations is
given by
\begin{equation}
\label{standardconsistent} \frac{A_T^2}{A_S^2} = -
\frac{1}{2}n_T\,,
\end{equation}
where $n_T \equiv d \ln A_T^2 /d \ln k$ represents the tilt of the
tensorial spectrum and $k$ is comoving wavenumber.

An identical relation holds in 4D scalar-tensor and other
generalized Einstein theories~\cite{tg}, and also in the RS
scenario~\cite{hl1,hl2}, and in a 5D brane-world model where the
radion field is stabilized~\cite{gklr}. Formally, the degeneracy
in the brane-world models arises because the function that
parametrizes the corrections to the gravitational wave amplitude
satisfies a particular first-order differential
equation~\cite{hl1,st}.

Given the potential importance of the consistency relation as a
way of reducing the number of independent inflationary parameters,
and of testing the inflationary scenario, it is important to
investigate whether the above degeneracy is lifted when IG effects
are included in the bulk action as a correction to the RS model.

\subsection{Scalar perturbations from IG brane inflation}

We will follow the approach used in Ref.~\cite{mwbh} to calculate
amplitude of scalar (density) perturbations due to slow-roll
inflation on the brane in RS gravity. This has recently been
applied to study the amplitude of scalar perturbations produced
during inflation on the brane for Gauss-Bonnet~\cite{GB} and
induced gravity corrections~\cite{pz,zc}.

To estimate the primordial density perturbation produced by
inflaton field fluctuations on the brane, to lowest order in a
slow-roll approximation, it is sufficient to use the amplitude of
quantum fluctuations of an effectively massless scalar field in an
unperturbed de Sitter spacetime. This gives the classic result
$\langle \delta\phi^2 \rangle = (H/2\pi)^2$ for modes exiting the
Hubble scale, with comoving wavenumber $k=aH$, and entering the
overdamped long-wavelength regime. The field fluctuation then
determines the spatial curvature perturbation on uniform-energy
density hypersurfaces, $\zeta=-H\delta\phi/\dot\phi$. Here and in
similar expressions in this section, equality is to be understood
as equality at the lowest order in the slow-roll approximation.

The curvature perturbation, $\zeta$, then remains constant for
adiabatic perturbations on large scales. This holds as a direct
consequence of the local conservation of energy-momentum on the
brane~\cite{wmll,lw} and, in particular, is independent of
gravitational field equations. Hence we can calculate $\zeta$ at
Hubble-exit during inflation and relate it directly to the
primordial density perturbation on large scales, as for instance
observed in the CMB anisotropies, so long as the perturbations are
strictly adiabatic. This is necessarily the case for slow-roll,
single field inflation on the brane so long as there are no
additional light scalar degrees of freedom such as might arise
from bulk metric perturbations in a brane-world scenario. This in
turn is true in the extreme slow-roll limit in single brane
Randall-Sundrum models and we expect it to remain true when
induced gravity corrections are included. There is no scalar
zero-mode and we have seen for the tensor perturbations that all
the normalizable massive modes have $m^2>{9\over4}H^2$ and remain
suppressed during inflation.

\begin{center}
\begin{figure}[h]
\begin{center}
\includegraphics[width=9cm]{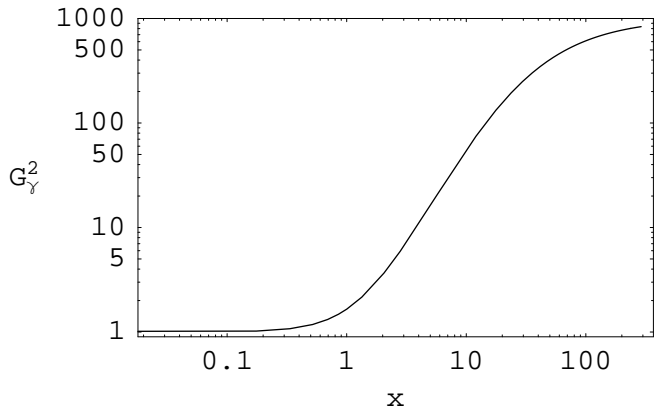}\\
\caption{The dimensionless amplitude $G^2_{\gamma}$ of density
perturbations relative to the 4D general relativity result,
plotted against the dimensionless energy scale of inflation,
$H/\mu$, for $\gamma=0.1$.}\label{F2}
\end{center}
\end{figure}
\end{center}

Consequently, the amplitude of a given mode that re-enters the
Hubble radius long after inflation is given by~\cite{llkcba}
\begin{equation}
A_S^2 = \frac{4}{25} \langle \zeta^2 \rangle = \frac{H^4}{25\pi^2
\dot{\varphi}^2} \,.
\end{equation}
During slow-roll inflation the scalar field equation of motion,
$\dot{\varphi} = -V'(\varphi)/3H$, implies that the amplitude of
scalar perturbations can be written as
\begin{equation}
\label{scalaramplitude} A_S^2 = \frac{9}{25 \pi^2}
\frac{H^6}{V'^2}\,.
\end{equation}
Using Eqs.~(\ref{Friedmann2}) and (\ref{kappa}), we write this as
 \begin{equation}
\label{AS} A_S^2 = \frac{\kappa_4^6 V^3}{75\pi^2 V'^2} \,
G^2_\gamma(H/\mu)\,,
 \end{equation}
where the correction to the standard 4D result is given by
\begin{equation} \label{gg}
G^2_\gamma(x)=\frac{x^6}{\left\{\gamma x^2
-2(1-\gamma)\left[1-\sqrt{1+x^2}\,\right]\right\}^3}\,.
\end{equation}
The behaviour of $G_\gamma^2(x)$ is illustrated in Fig.~\ref{F2}.
At low energies, $H/\mu\to0$, we have $G_\gamma^2(H/\mu)\to1$ and
we recover the standard 4D result. At higher energies we begin to
see an enhancement as in the RS case, but the enhancement of
perturbations is bounded when we include the IG correction, just
as we saw in the case of tensor perturbations. For $H\gg\mu$ and
$0<\gamma<1$ we obtain
\begin{equation} \label{ggh}
G^2_\gamma(H/\mu) \approx {1\over \gamma^3}\,.
\end{equation}

Equations~(\ref{Falpha}) and (\ref{ggh}) show that the scalars are
more strongly enhanced at high energies than the tensors, so that
the tensor/scalar ratio is suppressed at high energies relative to
the standard 4D result.

\subsection{Consistency relation}

To see whether the consistency relation (\ref{standardconsistent})
is broken by induced gravity, we need to calculate the
scale-dependence of the tensor spectrum,
\begin{equation}
n_T = \frac{d\ln A_T^2}{d\ln k} \,.
\end{equation}
We thus differentiate Eq.~(\ref{tensoramplitude}) with respect to
comoving wavenumber $k (\varphi) =a (\varphi) H (\varphi)$. In the
extreme slow-roll limit, variations in the Hubble parameter $H$
are negligible relative to changes in the scale factor and hence
the spectral index can be written as
\begin{equation}
\label{tensortilt} n_T =- \frac{d \ln  \left( HF_\gamma
\right)^{-2}}{d \ln H}\, \frac{d\ln H}{d\ln a}  \,.
\end{equation}

It is convenient to rewrite the scalar amplitude
(\ref{scalaramplitude}) in the form
\begin{equation}
\label{altAS} A_S^2 = - \frac{3}{25\pi} H^3 \frac{dH}{dV} \left(
\frac{d\ln H}{d\ln
    a} \right)^{-1} \,,
\end{equation}
where we have used the scalar field slow-roll equation
$3H\dot\varphi=-V'$, rewritten in the unusual form~\cite{GB}
\begin{equation}
\label{scalareom} V^{\prime2} = -3H^3 \frac{dV}{dH} \frac{d\ln
H}{d\ln a} \,.
\end{equation}
Then Eqs.~(\ref{tensoramplitude}), (\ref{tensortilt}) and
(\ref{altAS}) can be combined to give
\begin{equation}
\frac{A_T^2}{A_S^2}=-{Q \over 2}n_T \,,
\end{equation}
where
\begin{equation}
\label{defQ} Q = - \frac{\kappa_4^2}{3H} \frac{dV}{dH} F_\gamma^2
 \left[ \frac{d \ln (HF_\gamma)^{-2}}{d\ln H} \right]^{-1} .
\end{equation}

For the consistency relation (\ref{standardconsistent}) to hold,
we require $Q=1$. This amounts to a first-order differential
relation between $H(V)$ and $F_\gamma(H/\mu)$:
\begin{equation}
\frac{d}{dV} (3H^2) = \kappa_4^2 F_\gamma^2 \left( 1 + \frac{d\ln
F_\gamma}{d\ln H} \right)^{-1} \,.
\end{equation}
It is easy to see that this is satisfied at low energies, where
$F_\gamma\to1$ and $3H^2\to\kappa_4^2 V$, and we recover the
standard 4D case. It is far from obvious that this continues to
hold at higher energies. In fact one can show that the IG
brane-world correction to the tensor power spectrum,
Eq.~(\ref{F}), satisfies a first-order differential equation:
\begin{equation}
\label{importantode} \frac{d}{d \ln x} \left[ \ln
(xF_{\gamma})^{-2} \right] =
-2F^2_{\gamma}\left[\gamma+\frac{1-\gamma}
{\sqrt{1+x^2}}\right]\,.
\end{equation}
Combining the modified Friedmann equation (\ref{Friedmann2}) and
Eq.~(\ref{importantode}), it can then be shown that
Eq.~(\ref{defQ}) implies $Q=1$. Consequently, the brane gravity
model with IG correction (in the negative branch) does indeed
satisfy the consistency relation (\ref{standardconsistent}) for
any value of $H/\mu$. This extends the result originally found in
Ref.~\cite{hl1,hl2} for the RS where $\gamma=0$, to any IG
correction term with $0\leq\gamma<1$.

\section{Conclusions}

Brane-world inflation models allows one to explore some of the
cosmological implications of ideas arising from string and
M~theory. The extra-dimensional nature of gravity introduces new
features that need to be computed and then subjected to the
constraints from high-precision cosmological data. We have
determined the corrections to the standard results for tensor
perturbations that are generated during slow-roll inflation at
energies where brane effects become dominant. This has previously
been done for the Randall-Sundrum brane-world. We have introduced
an induced gravity term, as a correction to the gravitational
action. This correction leads to significant qualitative changes.

The 5D wave equation and its fundamental solutions are not changed
by the IG term. The spectrum contains a normalizable zero-mode and
a continuous tower of massive modes after a mass gap, $m>{3\over
2}H$, as in the RS case. The massive modes are not excited during
inflation, as in the RS case. However, the IG term changes the
boundary conditions at the brane, and therefore changes the
normalization of the zero-mode, as shown by Eq.~(\ref{psi0}). This
has the consequence of enhancing the amplitude of tensor
perturbations (gravitational waves) produced by de Sitter
inflation on the brane, Eq.(\ref{tensoramplitude}), relative to
the standard 4D result. Unlike the RS case, the relative
enhancement at high energies is bounded in the presence of a IG
correction term. However, as in the RS case, we find that the
standard consistency relation between the tensor-scalar ratio and
the tilt of the tensor spectrum remains unchanged.

\[ \]
{\bf Acknowledgements:}

We thank Lefteris Papantonopoulos for helpful discussions. MBL is
funded by the Spanish Ministry of Education, Culture and Sport
(MECD), and partly supported by DGICYT under Research Project
BMF2002~03758. RM is supported by PPARC and DW is supported by the
Royal Society.


\begin{thebibliography}{99}


\bibitem{review}
D. Langlois, arXiv:astro-ph/0301022; P. Brax and C. van de Bruck,
Class. Quantum Grav. {\bf 20}, R201 (2003) [arXiv:hep-th/0303095];
R. Maartens, Liv. Rev. Rel. {\bf 7} (2004) [arXiv:gr-qc/0312059];
P. Brax, C. van de Bruck, and A-C. Davis, arXiv:hep-th/0404011.

\bibitem{ch}
H.~Collins and B.~Holdom,
Phys.\ Rev.\ D {\bf 62}, 105009 (2000) [arXiv:hep-ph/0003173].

\bibitem{dgp}
G. R. Dvali, G. Gabadadze, and M. Porrati, Phys. Lett. B{\bf 485},
208 (2000) [arXiv:hep-th/0005016].

\bibitem{others}
Y. Shtanov, arXiv:hep-th/0005193; S.~Nojiri and S.~D.~Odintsov,
JHEP {\bf 0007}, 049 (2000) [arXiv:hep-th/0006232]; N.~J.~Kim,
H.~W.~Lee and Y.~S.~Myung,
Phys.\ Lett.\ B {\bf 504}, 323 (2001) [arXiv:hep-th/0101091];
G.~Kofinas,
JHEP {\bf 0108}, 034 (2001) [arXiv:hep-th/0108013].

\bibitem{deff}
C. Deffayet, Phys.\ Lett.\ B{\bf 502}, 199 (2001)
[arXiv:hep-th/0010186]; C.~Deffayet, G.~R.~Dvali and G.~Gabadadze,
Phys.\ Rev.\ D {\bf 65}, 044023 (2002) [arXiv:astro-ph/0105068];
C.~Deffayet, S.~J.~Landau, J.~Raux, M.~Zaldarriaga and P.~Astier,
Phys.\ Rev.\ D {\bf 66}, 024019 (2002) [arXiv:astro-ph/0201164];
V.~Sahni and Y.~Shtanov,
Int.\ J.\ Mod.\ Phys.\ D {\bf 11}, 1515 (2000)
[arXiv:gr-qc/0205111]; U.~Alam and V.~Sahni,
arXiv:astro-ph/0209443; R. G. Vishwakarma and P. Singh, Class.
Quantum Grav. {\bf 20}, 2033 (2003) [arXiv:astro-ph/0211285].

\bibitem{ktt}
E.~Kiritsis, N.~Tetradis and T.~N.~Tomaras,
JHEP {\bf 0203}, 019 (2002) [arXiv:hep-th/0202037].

\bibitem{mmt}
K. Maeda, S. Mizuno, and T. Torii, Phys.\ Rev.\ D {\bf 68}, 024033
(2003) [arXiv:gr-qc/0303039].

\bibitem{pz}
E. Papantonopoulos and V. Zamarias, arXiv:gr-qc/0403090.

\bibitem{zc}
H.~s.~Zhang and R.~G.~Cai,
arXiv:hep-th/0403234.

\bibitem{bw}
M. Bouhmadi-L\'opez and D. Wands, in preparation.

\bibitem{gs}
J. Garriga and M. Sasaki, Phys. Rev. D{\bf 62}, 043523 (2000)
[arXiv:hep-th/9912118].

\bibitem{lmw}
D. Langlois, R. Maartens, and D. Wands, Phys. Lett. B{\bf 489},
259 (2000) [arXiv:hep-th/0006007].

\bibitem{Bridgman:2001mc}
H.~A.~Bridgman, K.~A.~Malik and D.~Wands,
Phys.\ Rev.\ D {\bf 65}, 043502 (2002) [arXiv:astro-ph/0107245].

\bibitem{cdd} C. Charmousis and J-F. Dufaux, arXiv:hep-th/0311267.

\bibitem{GB} J.~F.~Dufaux, J.~E.~Lidsey, R.~Maartens and M.~Sami,
Phys. Rev. D, to appear, arXiv:hep-th/0404161.

\bibitem{2brane}
H. Davoudiasl, J. L. Hewett, and T. G. Rizzo, JHEP {\bf 03}, 034
(2003) [arXiv:hep-ph/0305086]; S. L. Dubovsky and M. V. Libanov,
JHEP {\bf 03}, 038 (2003) [arXiv:hep-th/0309131]; Y. Shtanov and
A. Viznyuk, arXiv:hep-th/0312261; M. N. Smolyakov,
arXiv:hep-th/0403034.

\bibitem{padilla}
A.~Padilla,
Class.\ Quant.\ Grav.\  {\bf 21}, 2899 (2004)
[arXiv:hep-th/0402079].

\bibitem{cks}
F. Cooper, A. Khare, and U. Sukhatme, Phys. Rep. {\bf 251}, 267
(1995) [arXiv:hep-th/9405029].

\bibitem{fk}
A. Frolov and L. Kofman, arXiv:hep-th/0209133.

\bibitem{grs}
D. S. Gorbunov, V. A. Rubakov, and S. M. Sibiryakov, J. High
Energy Phys. {\bf 10}, 015 (2001) [arXiv:hep-th/0108017];
T.~Kobayashi, H.~Kudoh, and T.~Tanaka, Phys.\ Rev.\ D{\bf 68},
044025 (2003) [arXiv:gr-qc/0305006].

\bibitem{llkcba}
J. E. Lidsey, A. R. Liddle, E. W. Kolb, E. J. Copeland, T.
Barreiro, and M. Abney, Rev.  Mod. Phys. {\bf 69}, 373 (1997).

\bibitem{liddle}
S. M. Leach and A. R. Liddle, Phys. Rev. D{\bf 68}, 123508 (2003)
[arXiv:astro-ph/0306305]; S. Tsujikawa and A. R. Liddle, JCAP {\bf
03}, 001 (2004) [arXiv:astro-ph/0312162]; S. Tsujikawa, M. Sami,
and R. Maartens, arXiv:astro-ph/0406078.

\bibitem{mwbh}
R. Maartens, D. Wands, B. A. Bassett, and I. P. C. Heard, Phys.
Rev. D{\bf 62}, 041301 (2000) [arXiv:hep-ph/9912464].

\bibitem{tg}
S. Tsujikawa and B. Gumjudpai, arXiv:astro-ph/0402185.

\bibitem{hl1}
G. Huey and J. E. Lidsey, Phys. Lett. B{\bf 514}, 217 (2001)
[arXiv:astro-ph/0104006].

\bibitem{hl2}
G. Huey and J. E. Lidsey, Phys. Rev. D{\bf 66}, 043514 (2002)
[arXiv:astro-ph/0205236].

\bibitem{gklr}
G. F. Giudice, E. W. Kolb, J. Lesgourgues, and A. Riotto, Phys.
Rev. D{\bf 66}, 083512 (2002) [arXiv:hep-ph/0207145].

\bibitem{st}
D. Seery and A. N. Taylor, arXiv:astro-ph/0309152; G. Calcagni,
JCAP {\bf 11}, 009 (2003) [arXiv:hep-ph/0310304].


\bibitem{wmll}
D. Wands, K. A. Malik, D. H. Lyth, and A. R. Liddle, Phys. Rev.
D{\bf 62}, 043527 (2000) [arXiv:astro-ph/0003278].

\bibitem{lw}
D.~H.~Lyth and D.~Wands,
Phys.\ Rev.\ D {\bf 68}, 103515 (2003) [arXiv:astro-ph/0306498].



\end{thebibliography}
\end{document}